\documentclass[a4paper,11pt]{article}
\usepackage{pos}
\usepackage{physics}
\usepackage{lineno}

\newcommand \LST{LST-1}
\newcommand \Fermi{$Fermi$-LAT}

\title{LST-1 observations of an enormous flare of BL~Lacertae in 2021}
 \ShortTitle{LST observations of BL~Lacertae flare in 2021}

\author*[a]{Seiya Nozaki}
\author[b]{Katsuaki Asano}
\author[c]{Juan Escudero}
\author[d]{Gabriel Emery}
\author[e]{Chaitanya Priyadarshi}

\affiliation[a]{Max-Planck-Institut für Physik, Föhringer Ring 6, 80805 München, Germany}
\affiliation[b]{Institute for Cosmic Ray Research, University of Tokyo, 5-1-5, Kashiwa-no-ha, Kashiwa, Chiba 277-8582, Japan}
\affiliation[c]{Instituto de Astrofísica de Andalucía-CSIC, Glorieta de la Astronomía s/n, 18008, Granada, Spain}
\affiliation[d]{Aix Marseille Univ, CNRS/IN2P3, CPPM, Marseille, France}
\affiliation[e]{Institut de Fisica d'Altes Energies (IFAE), The Barcelona Institute of Science and Technology, Campus UAB, 08193 Bellaterra (Barcelona), Spain}

\onbehalf{on behalf of the CTA-LST Project} 

\emailAdd{nozaki@mpp.mpg.de}

\abstract{
The first prototype of LST (LST-1) for the Cherenkov Telescope Array has been in commissioning phase since 2018 and already started scientific observations with the low energy threshold around a few tens of GeV.
In 2021, LST-1 observed BL Lac following the alerts based on multi-wavelength observations and detected prominent gamma-ray flares. In addition to the daily flux variability, LST-1 also detected sub-hour-scale intra-night variability reaching 3--4 times higher than the gamma-ray flux from the Crab Nebula above 100 GeV. 
In this proceeding, we will report the analysis results of LST-1 observations of BL Lac in 2021, especially focusing on flux variability. 
}

\ConferenceLogo{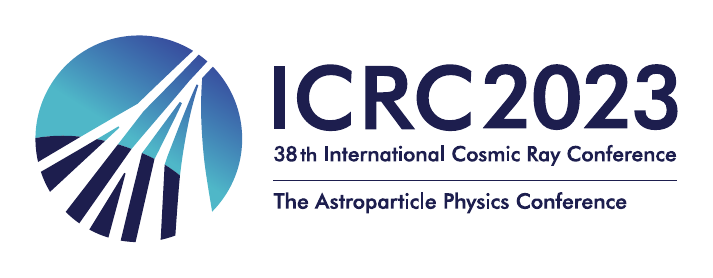}

\FullConference{%
38th International Cosmic Ray Conference (ICRC2023)\\
  26 July - 3 August, 2023\\
  Nagoya, Japan}

\begin{document}
\maketitle

\section{Introduction}
Blazars are a type of active galactic nuclei (AGN) characterized by the presence of collimated relativistic plasma jets oriented toward the Earth. The emission from blazars is characterized by a highly variable non-thermal electromagnetic spectrum from radio to very-high-energy gamma rays (VHE; E>$\sim20~\mathrm{GeV}$). The broadband SED has a two-hump structure. The lower-energy hump has a peak located from the radio to X-ray band and it is explained by the synchrotron emission from the accelerated leptons in the jet. 
On the other hand, the origin of the higher-energy hump located at gamma-ray bands is still under debate. Possible scenarios of the origin are inverse Compton scattering on low-energy photons emitted by synchrotron radiation and/or external photons (e.g. broad line region, dust torus).
Blazars with no or faint optical emission lines are classified as "BL Lac" type object. In addition, it is sub-classified based on the peak frequency of the synchrotron peak ($\nu_{s}$).

BL Lacertae (hereafter BL Lac) is a well-studied blazar located at redshift $z=0.069$~\citep{miller78}. 
BL~Lac is eponymous of the intermediate-synchrotron-peak "BL Lac" type object ($10^{14}\,\rm{Hz}<\nu_{s}<10^{15}\,\rm{Hz}$)~\citep{ackermann11}.
It is well known for the flux variability in various energy bands as described in~\citep{sahakyan22}. In the VHE gamma-ray band, BL Lac is only detected during the flaring state so far. After the first detection in VHE gamma-ray band (above 1~TeV) in 1998~\citep{neshpor01}, MAGIC and VERITAS detected multiple flares of BL Lac with various flux levels~\citep{magic07, arlen13, abeysekara18, magic19}. Some observations also detected intra-night flux variability. As an example, VERITAS detected a decay time of $13\pm4$~min in 2011~\citep{arlen13} and a rise and decay time of 2.3 hours and 36~minutes in 2016, respectively~\citep{abeysekara18}. Since 2019, BL Lac was relatively active in the gamma-ray band and VHE gamma-ray flares were detected by MAGIC several times~\citep{2019Atel_magic, 2020aAtel_magic, 2020bAtel_magic}.

Cherenkov Telescope Array\footnote{\url{https://www.cta-observatory.org/}} (CTA) will be a next-generation very-high-energy gamma-ray observatory. Three different sizes of telescopes are planned to be built to cover a wide energy range with an order of magnitude better sensitivity than the current generation of Cherenkov telescopes. The Large-Sized Telescope\footnote{\url{https://www.lst1.iac.es/}}, with a 23-m diameter mirror dish, is designed to detect (relatively) low-energy signals, upwards from a few tens of GeV. 
The first prototype of LST (LST-1) was inaugurated at the CTA northern site (La Palma, Spain) in October 2018 and it has been in the commissioning phase. In parallel with the commissioning tests, LST-1 already started to observe gamma-ray objects for scientific purposes.

In this contribution, we present the analysis results of the LST-1 observations of an enormous flare of BL Lac in 2021. 

\section{LST observations and analysis}
LST-1 started a campaign of BL Lac observations in July 2021 following the detection of the highest flux ever observed in the optical band~\citep{2021ATel_optical}. On July 11 (MJD 59406), LST-1 detected VHE gamma-ray signals from BL Lac despite under bad weather conditions~\citep{2021ATel_lst}. In August 2021, BL~Lac was still active and VHE gamma-ray signals were also detected by MAGIC telescopes~\citep{2021ATel_magic}. LST-1 continued its observation campaign until the mid of August 2021. 

The LST observations of BL Lac were performed during moonless time. We limited our observations to time windows with the source located at less than $\sim50$ degrees away from the zenith. It allows to take advantage of the LST-1 low-energy performance since low-energy photons are more absorbed by the thicker atmosphere crossed at lower altitudes. Each observation run takes 15--20 minutes. The total observation durations were 4.9 and 12.6 hours in July and August, respectively. 
We selected good-quality data based on the camera-averaged rate of pixel pulses with charge above 30~p.e as performed in~\citep{lst_performance23} representing the quality of the atmospheric conditions.
After the data selection, most of the LST-1 data taken in July 2021 were not selected. Thus, we only use the August datasets in this contribution and the duration of the selected August dataset is 9.8 hours. The observation of each night is summarized in Table~\ref{tab:bllac_observation}.
\begin{table*}[]
\caption{LST observation conditions.}
\label{tab:bllac_observation}
\centering
\begin{tabular}{ccccc} 
\hline \hline
Date & MJD & Observation time & $Zd$ range\\
& & [hours] & [degrees]\\ 
\hline
Aug 3 & 59428.95--59429.05 & 1.76 & 20--45\\
Aug 4 & 59429.94--59430.13 & 1.92 & 14--49\\
Aug 5 & 59431.10--59431.13 & 0.50 & 14--16\\
Aug 6 & 59431.92--59431.95 & 0.55 & 44--51\\
Aug 8 & 59434.20--59434.22 & 0.44 & 34--41\\
Aug 9 & 59434.99--59435.09 & 1.93 & 14--32\\
Aug 10 & 59436.04--59436.10 & 1.34 & 13--19\\
Aug 12 & 59438.12--59438.17 & 0.92 & 19--30\\
Aug 13 & 59439.03--59439.05 & 0.45 & 15--20\\
\hline
\end{tabular}
\end{table*}

The selected data were processed using the standard pipeline \texttt{cta-lstchain}\footnote{\url{https://github.com/cta-observatory/cta-lstchain}}~\citep{lst_performance_icrc2021, lstchain_v0.9.9}. The detail of the analysis procedures is described in~\citep{lst_performance23}. 
In this contribution, we performed the analysis using a likelihood technique\footnote{available in \texttt{cta-lstchain} v0.9.7+}, of which an earlier version is covered in~\citep{emery21}, to parameterize the data instead of the Hillas' parameters extraction used in~\citep{lst_performance23}. The method performs a fit of a space-time signal model at the waveform level. The choice was also made to perform a so called source-dependent analysis, with the assumption of the knowledge of the source position being used in the event reconstruction.
For the gamma-like event selection, we use $alpha$ (angle between shower axis and the line between the known source position and the image centroid) and $gammaness$ (the score indicating how likely it is that the primary particle is a gamma ray) obtained by machine learning. In addition, we apply an event cut of $intensity$ (the sum of the charges of the pixels which survive the image cleaning) above 50~photo-electrons to ensure the data quality.
The instrumental response function (IRF) of Cherenkov telescopes depends on the pointing direction of the telescope. Thus, we compute the IRF by the interpolation of ones obtained with the simulation data at different pointing directions.
The high level analysis was performed using \texttt{gammapy}\footnote{\url{https://github.com/gammapy/gammapy}}~\citep{gammapy17, gammapy_v0.20} to find the best spectral model using likelihood ratio test and compute the integrated gamma-ray flux.

\section{Multi-wavelength observations and analysis}
\subsection{Fermi-LAT}
The Large Area Telescope (LAT) on board the $Fermi$ Gamma-ray Space Telescope is a wide field-of-view pair-conversion telescope covering the energy range from below 20 MeV to more than 300 GeV~\citep{atwood09}. It has a wide field of view so that the entire sky is scanned every three hours for the standard survey mode. We analyzed \Fermi{} data between July 29, 2021 (MJD=59424) and August 15, 2021 (MJD=59441) covering the \LST{} observation periods in August 2021. We performed the binned likelihood analysis using \texttt{fermipy}\footnote{\url{https://github.com/fermiPy/fermipy}}.
The analysis settings follow the recommendation for the Pass 8 data analysis\footnote{\url{https://fermi.gsfc.nasa.gov/ssc/data/analysis/documentation/Pass8_usage.html}}. We analyzed the data in 24-hours and 12-hours bins.

\subsection{Swift-UVOT, XRT}
The Neil Gehrels $Swift$ Observatory is a multiwavelength mission for Gamma-Ray Burst astronomy and has been operational since 2004~\citep{gehrels04}. $Swift$ carried out 13 observations of BL~Lac around the LST-1 observation campaign in August 2021. In this proceeding, we used the data obtained by the X-ray Telescope (XRT) \citep{burrows04} (0.2–10.0 keV) and the
Ultraviolet/Optical Telescope (UVOT)~\citep{roming05} (170–600 nm) onboard the $Swift$ satellite. XRT data were processed using an online XRT product generator\footnote{\url{https://www.swift.ac.uk/user_objects/}} and UVOT data were analyzed using the python wrapper tool including the official UVOT analysis pipeline\footnote{\url{https://www.swift.ac.uk/analysis/uvot/}}\footnote{\url{https://github.com/KarlenS/swift-uvot-analysis-tools}}

\section{Results}
\subsection{VHE gamma-ray signal detection}
\begin{figure}
  \begin{center}
  \includegraphics[width=.6\textwidth]{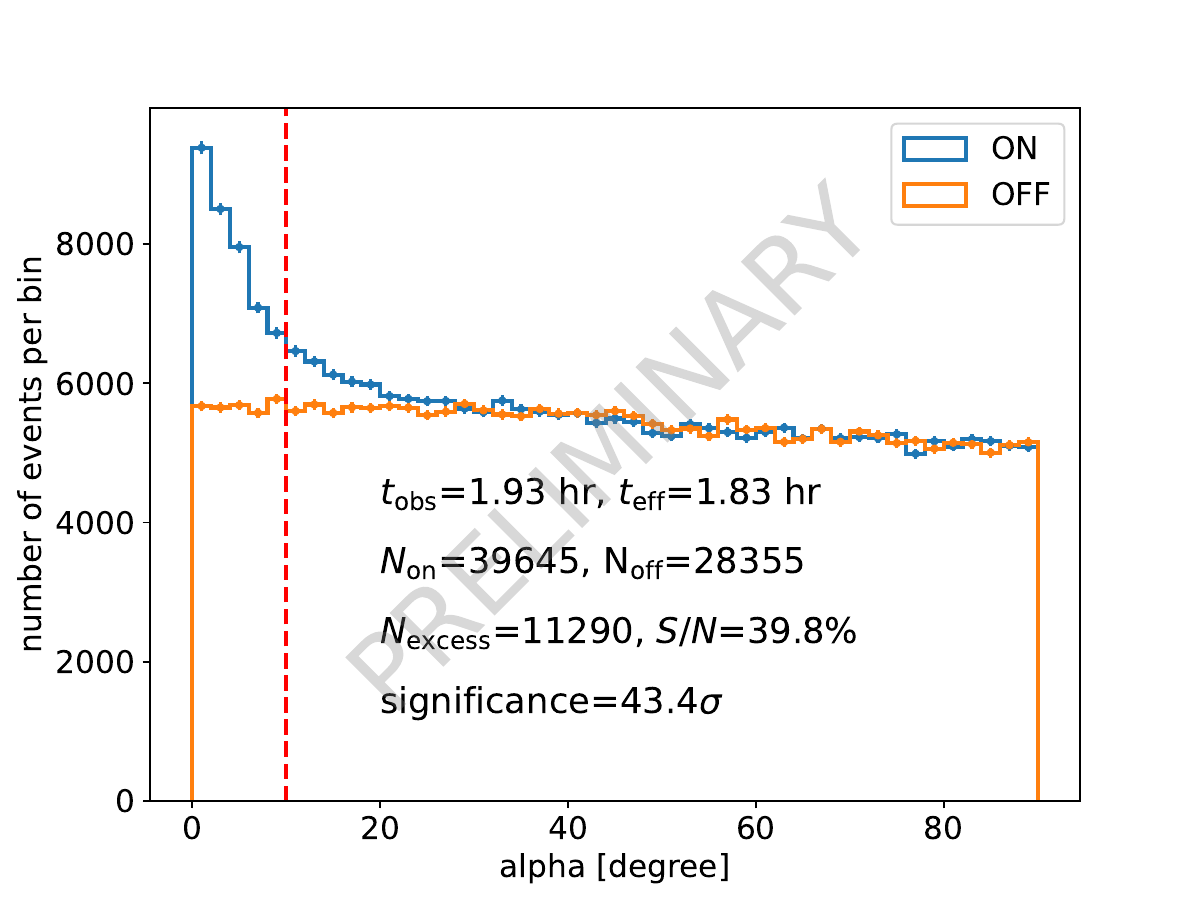}
  \caption{Distributions of $alpha$ of the LST observation on Aug 9. Blue and orange histograms correspond to the distribution of ON and OFF events, respectively. Each error bar shows the statistical uncertainty. Here, we apply event cuts of $intensity>50~\rm{p.e.}$ and $gammaness>0.9$. To compute the statistics in the figure, a cut of $alpha < 10\,\mathrm{degrees}$ (red-dashed line) is also applied.}
  \label{fig:alpha}
  \end{center}
\end{figure}

Fig.~\ref{fig:alpha} shows the distribution of $alpha$ from the expected source position (ON) and another position without gamma-ray sources (OFF) on Aug 9.
%
The VHE gamma-ray signals are clearly detected with a significance of $43.4\sigma$.
The background level at a larger $alpha$ region is consistent between ON and OFF events. Even though the difference in the background level is small, it can significantly affect the results at the lowest energy range since the background rejection power of the single telescope data analysis is worse than the stereo data analysis below $\sim100~\mathrm{GeV}$ as seen in Fig.~16 of \citep{lst_performance23}. For this flare, the signal-to-noise ratio is high even for the low-energy events so that the results are less affected by the background normalization factors than other gamma-ray sources.

\subsection{VHE gamma-ray light curve}
\begin{figure}
  \begin{center}
  \includegraphics[width=.6\textwidth]{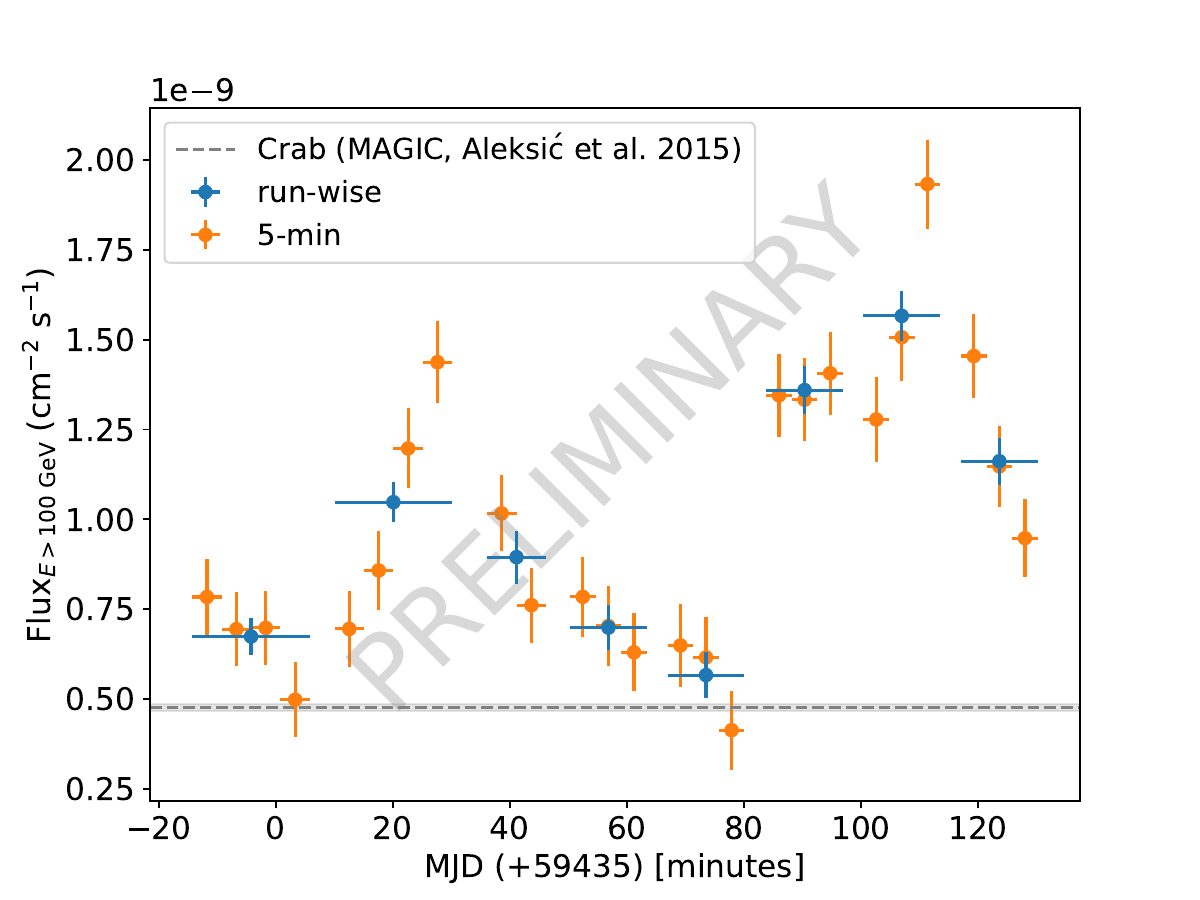}
  \caption{Intra-night light curve (>100 GeV) observed by the LST-1 on Aug 9. Blue and orange points correspond to run-wise and 5-min duration. The error bars on the flux level show statistical uncertainties. The gray line shows the integral flux of the Crab Nebula obtained by the MAGIC~\citep{magic_crab} as a reference. 
  }
  \label{fig:sed_lc}
  \end{center}
\end{figure}

Fig.~\ref{fig:sed_lc} shows the intra-night light curve above 100~GeV on Aug 9. The flux level was variable during this night and reached 3--4 times higher than that of the Crab Nebula (Crab Unit; C.U.) at maximum. Two peaks can be seen in the intra-night light curve, both with a rise and decay time scale around 10–20 minutes.
%

\subsection{Multi-wavelength light curve}
\begin{figure*}
  \begin{center}
  \includegraphics[width=1.0\textwidth]{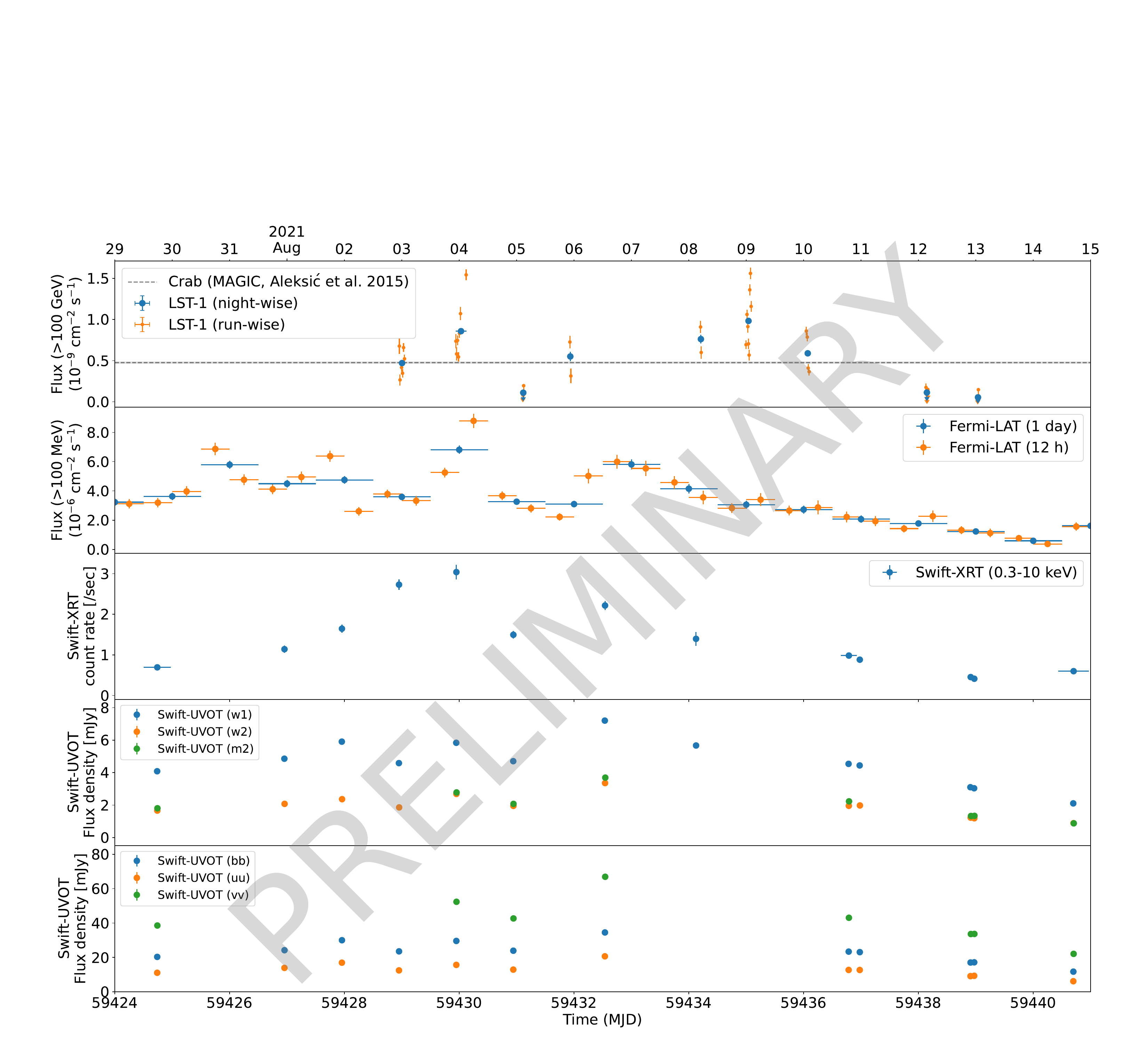}
  \caption{Multi-wavelength light curves of the BL Lacertae between MJD=59424 and 59441  (VHE gamma-ray: LST-1 (>100 GeV), high-energy gamma-ray: Fermi-LAT (> 100 MeV), X-ray: Swift-XRT (0.3--10 keV), Ultraviolet and Optical: Swift-UVOT). For LST-1, night-wise (blue) and run-wise (orange) light curves are shown. Gray dashed line shows the integral flux of the Crab Nebula~\citep{magic_crab} as shown in Fig.~\ref{fig:sed_lc}. For Fermi-LAT, 1-day (blue) and 12-day (orange) light curves are shown.}
  \label{fig:mwl_lc}
  \end{center}
\end{figure*}

Fig.~\ref{fig:mwl_lc} shows the multi-wavelength light curve around the LST-1 observation period. 
In the night-wise LST light curve, we can see the flux variation (<0.1--2 C.U.) on a long time scale with large intra-night variabilities. In the \Fermi{} light curve, multiple peaks were detected during this period and the highest flux was observed around Aug 4 when LST-1 also detected high flux ($\sim3$~C.U.) in the run-wise light curve. Around this peak, Swift-XRT also detected the highest count rates reaching around three times higher than the lowest rates during this period. After all of the bands showed an increase in the flux again around Aug 6--7, it gradually decreased except for the LST-1 data points observed on Aug 9.

%

\section{Summary}
LST-1 performed the observation of BL Lac flare in 2021. During the observation campaign, VHE gamma-ray signals were detected on most nights and the detection significance was 43.4$\sigma$ on Aug 9 with the high signal-to-noise ratio. 
We have observed the intra-night variability with the flux level of 3--4 C.U. at the peak. This variability had a sub-hour scale double-peak structure. 
The LST-1 night-wise light curve also shows the daily-scale flux variability  (<0.1--2 C.U.). \Fermi{} and Swift-XRT show the highest flux around Aug 4 when LST-1 also detected a high flux of $\sim3$~C.U. in the run-wise light curve. 
The detailed discussion and interpretation will be described in a future paper.

\section*{Acknowledgments} 
We gratefully acknowledge financial support from the following agencies and organisations:
{\scriptsize
Conselho Nacional de Desenvolvimento Cient\'{\i}fico e Tecnol\'{o}gico (CNPq), Funda\c{c}\~{a}o de Amparo \`{a} Pesquisa do Estado do Rio de Janeiro (FAPERJ), Funda\c{c}\~{a}o de Amparo \`{a} Pesquisa do Estado de S\~{a}o Paulo (FAPESP), Funda\c{c}\~{a}o de Apoio \`{a} Ci\^encia, Tecnologia e Inova\c{c}\~{a}o do Paran\'a - Funda\c{c}\~{a}o Arauc\'aria, Ministry of Science, Technology, Innovations and Communications (MCTIC), Brasil;
Ministry of Education and Science, National RI Roadmap Project DO1-153/28.08.2018, Bulgaria;
Croatian Science Foundation, Rudjer Boskovic Institute, University of Osijek, University of Rijeka, University of Split, Faculty of Electrical Engineering, Mechanical Engineering and Naval Architecture, University of Zagreb, Faculty of Electrical Engineering and Computing, Croatia;
Ministry of Education, Youth and Sports, MEYS  LM2015046, LM2018105, LTT17006, EU/MEYS CZ.02.1.01/0.0/0.0/16\_013/0001403, CZ.02.1.01/0.0/0.0/18\_046/0016007 and CZ.02.1.01/0.0/0.0/16\_019/0000754, Czech Republic; 
CNRS-IN2P3, the French Programme d’investissements d’avenir and the Enigmass Labex, 
This work has been done thanks to the facilities offered by the Univ. Savoie Mont Blanc - CNRS/IN2P3 MUST computing center, France;
Max Planck Society, German Bundesministerium f{\"u}r Bildung und Forschung (Verbundforschung / ErUM), Deutsche Forschungsgemeinschaft (SFBs 876 and 1491), Germany;
Istituto Nazionale di Astrofisica (INAF), Istituto Nazionale di Fisica Nucleare (INFN), Italian Ministry for University and Research (MUR);
ICRR, University of Tokyo, JSPS, MEXT, Japan;
JST SPRING - JPMJSP2108;
Narodowe Centrum Nauki, grant number 2019/34/E/ST9/00224, Poland;
The Spanish groups acknowledge the Spanish Ministry of Science and Innovation and the Spanish Research State Agency (AEI) through the government budget lines PGE2021/28.06.000X.411.01, PGE2022/28.06.000X.411.01 and PGE2022/28.06.000X.711.04, and grants PID2022-139117NB-C44, PID2019-104114RB-C31,  PID2019-107847RB-C44, PID2019-104114RB-C32, PID2019-105510GB-C31, PID2019-104114RB-C33, PID2019-107847RB-C41, PID2019-107847RB-C43, PID2019-107847RB-C42, PID2019-107988GB-C22, PID2021-124581OB-I00, PID2021-125331NB-I00; the ``Centro de Excelencia Severo Ochoa" program through grants no. CEX2019-000920-S, CEX2020-001007-S, CEX2021-001131-S; the ``Unidad de Excelencia Mar\'ia de Maeztu" program through grants no. CEX2019-000918-M, CEX2020-001058-M; the ``Ram\'on y Cajal" program through grants RYC2021-032552-I, RYC2021-032991-I, RYC2020-028639-I and RYC-2017-22665; the ``Juan de la Cierva-Incorporaci\'on" program through grants no. IJC2018-037195-I, IJC2019-040315-I. They also acknowledge the ``Atracción de Talento" program of Comunidad de Madrid through grant no. 2019-T2/TIC-12900; the project ``Tecnologi\'as avanzadas para la exploracio\'n del universo y sus componentes" (PR47/21 TAU), funded by Comunidad de Madrid, by the Recovery, Transformation and Resilience Plan from the Spanish State, and by NextGenerationEU from the European Union through the Recovery and Resilience Facility; the La Caixa Banking Foundation, grant no. LCF/BQ/PI21/11830030; the ``Programa Operativo" FEDER 2014-2020, Consejer\'ia de Econom\'ia y Conocimiento de la Junta de Andaluc\'ia (Ref. 1257737), PAIDI 2020 (Ref. P18-FR-1580) and Universidad de Ja\'en; ``Programa Operativo de Crecimiento Inteligente" FEDER 2014-2020 (Ref.~ESFRI-2017-IAC-12), Ministerio de Ciencia e Innovaci\'on, 15\% co-financed by Consejer\'ia de Econom\'ia, Industria, Comercio y Conocimiento del Gobierno de Canarias; the ``CERCA" program and the grant 2021SGR00426, both funded by the Generalitat de Catalunya; and the European Union's ``Horizon 2020" GA:824064 and NextGenerationEU (PRTR-C17.I1).
State Secretariat for Education, Research and Innovation (SERI) and Swiss National Science Foundation (SNSF), Switzerland;
The research leading to these results has received funding from the European Union's Seventh Framework Programme (FP7/2007-2013) under grant agreements No~262053 and No~317446;
This project is receiving funding from the European Union's Horizon 2020 research and innovation programs under agreement No~676134;
ESCAPE - The European Science Cluster of Astronomy \& Particle Physics ESFRI Research Infrastructures has received funding from the European Union’s Horizon 2020 research and innovation programme under Grant Agreement no. 824064.
}

\clearpage
\section*{Full Author List: CTA-LST project}
%
%



\tiny{\noindent
K. Abe$^{1}$,
S. Abe$^{2}$,
A. Aguasca-Cabot$^{3}$,
I. Agudo$^{4}$,
N. Alvarez Crespo$^{5}$,
L. A. Antonelli$^{6}$,
C. Aramo$^{7}$,
A. Arbet-Engels$^{8}$,
C.  Arcaro$^{9}$,
M.  Artero$^{10}$
% ...
}\\

\tiny{\noindent
$^{1}$Department of Physics, Tokai University.
$^{2}$Institute for Cosmic Ray Research, University of Tokyo.
$^{3}$Departament de Física Quàntica i Astrofísica, Institut de Ciències del Cosmos, Universitat de Barcelona, IEEC-UB.
$^{4}$Instituto de Astrofísica de Andalucía-CSIC.
$^{5}$EMFTEL department and IPARCOS, Universidad Complutense de Madrid.
$^{6}$INAF - Osservatorio Astronomico di Roma.
$^{7}$INFN Sezione di Napoli.
$^{8}$Max-Planck-Institut für Physik.
$^{9}$INFN Sezione di Padova and Università degli Studi di Padova.
$^{10}$Institut de Fisica d'Altes Energies (IFAE), The Barcelona Institute of Science and Technology.
% ...
}

\end{document}